\documentclass[twocolumn,
showpacs,
preprintnumbers,
amsmath,
amssymb,
amsfonts,
final,
a4paper,
aps,
floatfix,
citeautoscript,
footinbib,
prb,
superscriptaddress]{revtex4}
\usepackage{times}
\usepackage{graphicx}
\usepackage[latin1]{inputenc}
\usepackage{mathrsfs}

\begin{document}
\title{Haldane charge conjecture in one-dimensional multicomponent
 fermionic cold atoms}
\author{H. Nonne}
\affiliation{Laboratoire de Physique Th\'eorique et
Mod\'elisation, CNRS UMR 8089,
Universit\'e de Cergy-Pontoise, Site de Saint-Martin,
F-95300 Cergy-Pontoise Cedex, France}
\author{P. Lecheminant}
\affiliation{Laboratoire de Physique Th\'eorique et
Mod\'elisation, CNRS UMR 8089,
Universit\'e de Cergy-Pontoise, Site de Saint-Martin,
F-95300 Cergy-Pontoise Cedex, France}
\author{S.\ Capponi} \affiliation{Laboratoire de
Physique Th\'eorique, CNRS UMR 5152, 
Universit\'e Paul Sabatier, F-31062 Toulouse, France.}
\author{G. Roux}
\affiliation{Laboratoire de Physique Th\'eorique et Mod\`eles Statistiques,
 Universit\'e Paris-Sud, CNRS UMR~8626, 91405 Orsay, France}
\author{E. Boulat}
\affiliation{Laboratoire Mat\'eriaux et Ph\'enom\`enes Quantiques,
CNRS UMR 7162, 
Universit\'e Paris Diderot, 
75013 Paris, France}

\date{\today}

\begin{abstract}
  A Haldane conjecture is revealed for spin-singlet charge modes in
  $2N$-component fermionic cold atoms loaded into a one-dimensional
  optical lattice. By means of a low-energy approach and DMRG
  calculations, we show the emergence of gapless and gapped phases
  depending on the parity of $N$ for attractive interactions at
  half-filling. The analogue of the Haldane phase of the spin-1
  Heisenberg chain is stabilized for $N=2$ with non-local string
  charge correlation, and pseudo-spin 1/2 edge states. At the heart of
  this even-odd behavior is the existence of a spin-singlet
  pseudo-spin $N/2$ operator which governs the low-energy properties
  of the model for attractive interactions and gives rise to the
  Haldane physics.
\end{abstract}

\maketitle

\section{Introduction}

One of the major advances in the understanding of low-dimensional
strongly correlated systems has been the so-called Haldane conjecture.
In 1983, Haldane argued that the spin-$S$ Heisenberg chain displays
striking different properties depending on the parity of
2$S$~\cite{haldane}. While half-integer Heisenberg spin chains have a
gapless behavior, a finite gap from the singlet ground state to the
first triplet excited states is found when 2$S$ is even. The Haldane
conjecture is now well understood and has been confirmed
experimentally and numerically. On top of the existence of a gap, the
spin-1 phase (the Haldane phase) has remarkable exotic properties.
This phase displays non-local string long-range ordering which
corresponds to the presence of a hidden N\'eel antiferromagnetic
order~\cite{dennijs}.  One of the most remarkable properties of the
Haldane phase is the liberation of fractional spin-1/2 edge states
when the chain is doped by non-magnetic
impurities~\cite{hagiwara}. The possibility of a similar hidden order
has recently been proposed in a different context, by studying the
one-dimensional extended Bose-Hubbard model~\cite{Berg2008}.

In this Rapid Communication, we will reveal a Haldane conjecture for
\textit{spin-singlet} modes in a $2N$-component fermionic chain at
half-filling and for \textit{attractive} interactions, with the
emergence of gapless and gapped phases depending on the parity of
$N$. The analog of the Haldane phase is stabilized for even $N$ with
all its well-known properties, while a gapless behavior occurs when
$N$ is odd. The Haldane physics with the alternating gapped or gapless
behavior thus translates here directly into an insulating or metallic
behavior depending on the parity of $N$. To exhibit this even-odd
scenario, we will consider cold fermionic atoms with half-integer
hyperfine spin $F = N -1/2$ at half-filling ($N$ atoms per site)
loaded into a one-dimensional optical lattice.  Due to Pauli's
principle, low-energy s-wave scattering processes of spin-$F$
fermionic atoms are allowed in the even total spin $J=0,2,\ldots,
2N-2$ channels, so that the effective Hamiltonian with contact
interactions reads as follows~\cite{ho}:

\begin{eqnarray}
{\cal H}
&=& -t \sum_{i,\alpha} \left[c^{\dagger}_{\alpha,i}
c_{\alpha, i+1} + {\rm H.c.} \right]
- \mu \sum_{i,\alpha} c^{\dagger}_{\alpha,i} c_{\alpha,i}
\nonumber \\
&+& \sum_{i,J} U_J \sum_{M=-J}^{J}
P_{JM,i}^{\dagger} P_{JM,i},
\label{hubbardSgen}
\end{eqnarray}
where $c^{\dagger}_{\alpha,i}$ is the fermion creation operator
corresponding to the $2N$ hyperfine states ($\alpha = 1,\ldots, 2N$) at
the $i^{\text{th}}$ site of the optical lattice. The pairing operators in
Eq.~(\ref{hubbardSgen}) are defined through the Clebsch-Gordan
coefficients for spin-$F$ fermions: $P^{\dagger}_{JM,i} = \sum_{\alpha
  \beta} \langle{JM|F,F;\alpha \beta}\rangle c^{\dagger}_{\alpha,i}
c^{\dagger}_{\beta,i}$. In the general spin-$F$ case, there are $N$
couplings constants $U_J$ in model~(\ref{hubbardSgen}) which are
related to the $N$ possible two-body scattering lengths of the
problem. In the following, we will consider a simplified version of
model~(\ref{hubbardSgen}) for $N\geq 2$ to reveal explicitly the Haldane
charge conjecture. By fine-tuning the different scattering lengths in
channel $J \ge 2$, we will investigate model (\ref{hubbardSgen}) with
$U_2 = ... = U_{2N-2}$:
\begin{eqnarray}
{\cal H} &=& -t \sum_{i,\alpha} [c^{\dagger}_{\alpha,i} c_{\alpha,i+1} +
{\rm H.c.} ]
- \mu \sum_i n_i
\nonumber \\
&+& \frac{U}{2} \sum_i n_i^2 + V \sum_i P^{\dagger}_{00,i}
P_{00,i},
\label{hubbardS}
\end{eqnarray}
with $U= 2 U_2$, $V = U_0 - U_2$, and $n_i = \sum_{\alpha}
n_{\alpha,i} = \sum_{\alpha} c^{\dagger}_{\alpha,i} c_{\alpha,i}$ is
the density at site $i$. In Eq.~(\ref{hubbardS}), the singlet BCS
pairing operator for spin-$F$ fermions is $ \sqrt{2N} P^{\dagger}_{00,i}
= \sum_{\alpha \beta} c^{\dagger}_{\alpha,i} {\cal J}_{\alpha \beta}
c^{\dagger}_{\beta,i} = - \sum_{\alpha} \left(-1\right)^{\alpha}
c^{\dagger}_{\alpha,i} c^{\dagger}_{2N+1-\alpha,i}$, where the matrix
${\cal J}$ is a $2N \times 2N$ antisymmetric matrix with ${\cal J}^2 =
- I$. When $V=0$ ($U_0=U_2$), model (\ref{hubbardS}) is nothing but
the Hubbard model for $2N$-component fermions with a U(2$N$)$=$ U(1)
$\times$ SU(2$N$) invariance. This symmetry is broken down to U(1)
$\times$ Sp(2$N$) when $V\ne 0$ \cite{sachdev,wuzhang}.  In the
special $N=2$ case, i.e. $F=3/2$, there is no fine-tuning and models
(\ref{hubbardSgen}) and (\ref{hubbardS}) have an exact U(1) $\times$
SO(5) symmetry [Sp(4) $\sim$ SO(5)] \cite{zhang}. The zero-temperature
phase diagram of model (\ref{hubbardS}) away from half-filling has
been recently investigated by means of a low-energy approach
\cite{Lecheminant2005,Wu2005} and large scale numerical calculations
\cite{sylvainMS} for $F=3/2$. In this respect, the physics of $F>1/2$
fermions is richer than in the standard spin-1/2 Hubbard chain
\cite{bookboso,giamarchi}.  In particular, for $U < V < 0$ and at
sufficiently low density, the leading superconducting instability is
of a molecular type with charge
$2Ne$~\cite{Lecheminant2005,Wu2005,sylvainMS}. In this Rapid
Communication, we will
show by means of a low-energy approach and density matrix
renormalization group (DMRG) calculations \cite{DMRG} that a Haldane
conjecture for {\it spin-singlet} charge modes emerges in model
(\ref{hubbardS}) at half-filling depending on the parity of $N$.  In
the $N=1$ case, it is well-known that the half-filled SU(2) Hubbard
chain displays a critical phase for attractive interaction.  The
analog of the Haldane phase of the spin-1 Heisenberg chain occurs
for $N=2$ and attractive interactions.

\section{Strong-coupling argument}  

We first give a simple physical
explanation of the emergence of the Haldane conjecture for charge
degrees of freedom. It stems from the existence of a pseudo-spin
operator which carries charge: ${\cal S}^{+}_i = \sqrt{N/2}
P^{\dagger}_{00,i}$ and ${\cal S}^{z}_i = ( n_i - N)/2$. This operator
is a Sp(2$N$) spin-singlet which is the generalization of the
$\eta$-pairing operator introduced by Yang for the half-filled
spin-1/2 (i.e. $N=1$) Hubbard model~\cite{yang}.  It is easy to
observe that ${\vec {\cal S}}_i$ satisfies the SU(2) commutation
relations and generates a higher SU(2) $\times$ Sp(2$N$) symmetry at
half-filling along a very special line $V = N U$.  The existence of
such an extended SU(2) symmetry in the charge sector for $N=2$ has
been first noticed in Ref.~\onlinecite{zhang}.  In the general $N$
case, one simple way to observe the emergence of this symmetry for $V
= N U$ is to rewrite the Hamiltonian (\ref{hubbardS}) in absence of
the hopping term [$\mu = U (N+1)$]: ${\cal H}(t=0) = 2U \sum_i
\left({\vec {\cal S}}^2_i - N(N+2)/4 \right)$. On top of the Sp(2$N$)
symmetry, we thus deduce the existence of an extended SU(2) symmetry
in the charge sector; moreover, for a strong attractive $U$, the
pseudo-spin ${\vec {\cal S}}_i$ is a spin-$N$/2 operator, which acts on
the degenerate low-lying even occupied states $({\cal
  S}_i^+)^k\left|{\emptyset}\right\rangle$ \cite{footnote}, which
one sketches here for $N=1,2$:
\begin{center}
\includegraphics{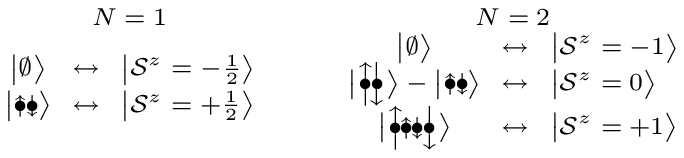}
\end{center}
The next step of the approach is to derive an effective Hamiltonian in
the strong coupling regime $|U| \gg t$. At second order of
perturbation theory and after a simple gauge transformation that changes the sign of the transverse
exchange,  we find a spin-$N$/2 antiferromagnetic SU(2)
Heisenberg chain: ${\cal H}_{\rm eff} = J \sum_i {\vec {\cal S}}_i
\cdot {\vec {\cal S}}_{i+1}$ with $J = 4 t^2/[N(2N+1)|U|]$. 
The
Haldane conjecture for model~(\ref{hubbardS}) with attractive
interactions thus becomes clear within this strong-coupling argument.
When we deviate from the $V = N U$ line, the SU(2) charge symmetry is
broken down to U(1) and in the strong-coupling regime the lowest
correction is a single-ion anisotropy $D \sum_i ({\cal S}^{z}_i)^2$
(with $D= 2[U-V/N]$).  The phase diagram of the resulting
model for general $N$ is known from the work of Schulz~\cite{schulz}.
For even $N$, on top of the Haldane phase, N\'eel and large-$D$ singlet
gapful phases appear while gapless ($XY$) and gapful (Ising) phases are
stabilized for odd $N$ in the vicinity of the SU(2) line.  

We now turn
to low-energy and numerical approaches to investigate the strong-weak
coupling cross-over and the determination of the physical properties
of the phases in the vicinity of the $V = N U$ line.

\begin{figure}[t]
\centering
\includegraphics[width=0.6\columnwidth,clip]{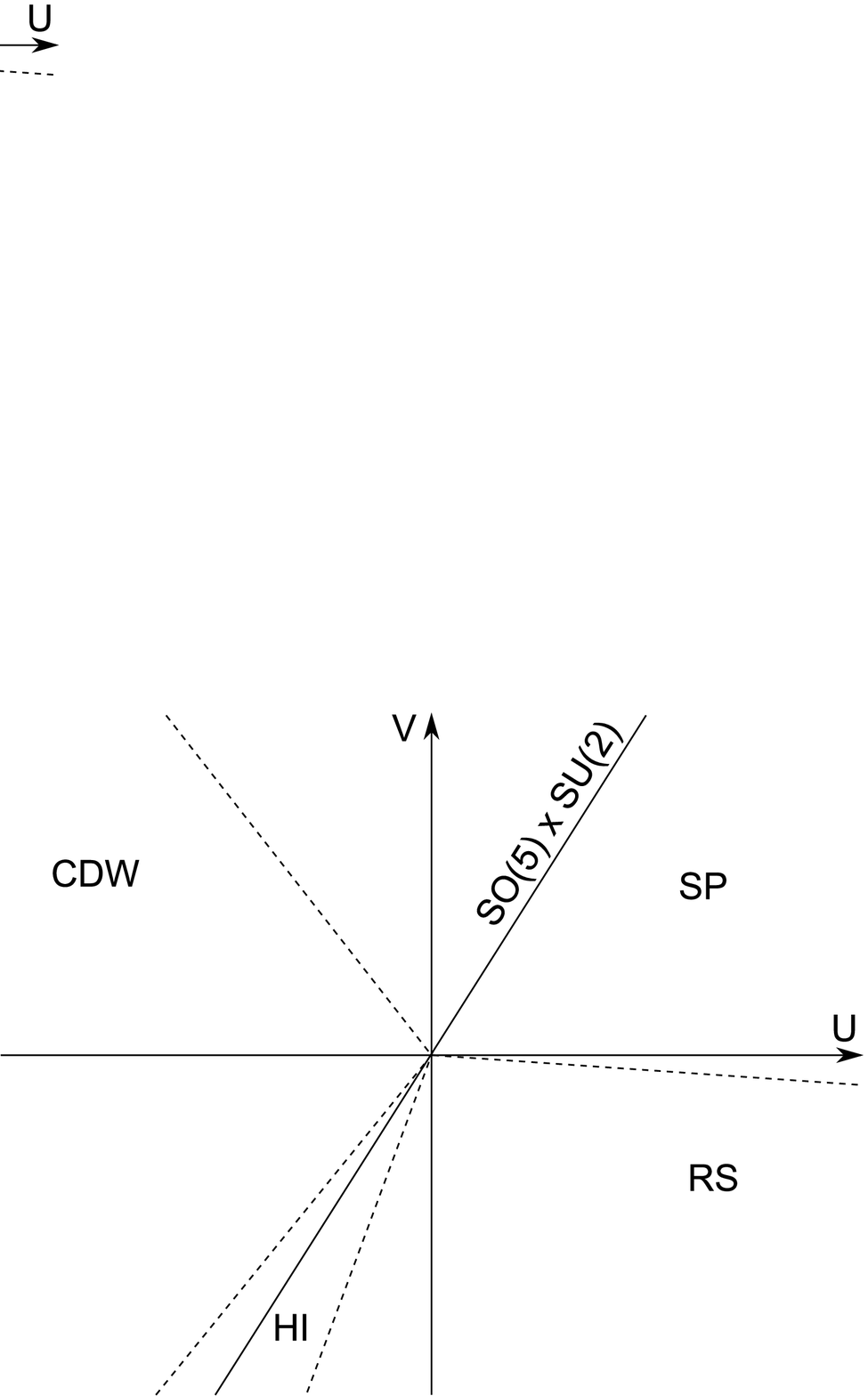}
\caption{Phase diagram obtained by the low-energy approach in the
  $N=2$ case (see text for definitions); the dotted lines stand for
  (second-order) quantum phase transitions.}
\label{fig1}
\end{figure}

\section{Low-energy approach}  We study here the low-energy
approach in the simplest $F=3/2$ case with the emergence of the
striking properties of a Haldane insulating (HI) phase.  The general
$N$ case is highly technical and will be presented elsewhere.  The
low-energy procedure for $F=3/2$ cold fermions has already been
presented away from
half-filling~\cite{Lecheminant2005,Wu2005,controzzi}.  In the
half-filled case, in sharp contrast to the $F=1/2$ case, there is no
spin-charge separation for $F > 1/2$ since an umklapp process couples
these degrees of freedom \cite{assaraf}.  The exact U(1) $\times$
SO(5) continuous symmetry of model (\ref{hubbardS}) is hidden in the
bosonization description. However, it becomes explicit by a
refermionization procedure as in the two-leg spin ladder
\cite{bookboso}. To this end, we introduce eight right and left moving
real (Majorana) fermions $\xi_{R,L}^{A}, A=1,\ldots,8$. The two
Majorana fermions $\xi^{7,8}$ account for the U(1) charge symmetry,
the five Majorana fermions $\xi^{1,\ldots,5}$ generate the SO(5) spin
rotational symmetry whereas the last one $\xi^{6}$ describes an
internal discrete $\mathbb{Z}_2$ symmetry
($c_{1(4),i}\rightarrow ic_{1(4),i}$, $c_{2(3),i}\rightarrow -ic_{2(3),i}$)
of model
(\ref{hubbardS}). Within this description, the interacting part of
the low-energy Hamiltonian for the spin-3/2 model (\ref{hubbardS}) at
half-filling reads as follows:
\begin{eqnarray}
{\cal H}_{\rm int} &=&
\frac{g_1}{2} \; \bigg(\sum_{a=1}^{5} \xi_R^{a} \xi_L^{a} \bigg)^2
+ g_2 \; \xi_R^{6} \xi_L^{6}
\sum_{a=1}^{5} \xi_R^{a} \xi_L^{a}
\nonumber \\
&+& \frac{g_3}{2} \; \left(\xi_R^{7} \xi_L^{7} 
+ \xi_R^{8} \xi_L^{8} \right)^2
\label{majofthalfilling}\\
&+& \left( \xi_R^{7} \xi_L^{7} + \xi_R^{8} \xi_L^{8} \right)
\bigg( g_4 \;
\sum_{a=1}^{5} \xi_R^{a} \xi_L^{a}
+ 
g_5 \; \xi_R^{6} \xi_L^{6} \bigg),
\nonumber 
\end{eqnarray}
with $g_{1,2} = -a_0 \left( U \pm V \right), g_3 = a_0 \left( 3 U +
  V\right), g_4 = a_0 U$, and $g_5 = a_0\left( U + 2 V\right)$. 
The zero-temperature
phase diagram of model (\ref{majofthalfilling}) can then be derived by
means of a one-loop renormalization group (RG) approach. By
neglecting the velocity anisotropy, we find the one-loop RG equations:
\begin{eqnarray}
{\dot g}_1 &=& 3 g_1^2 + g_2^2 + 2 g_4^2, 
\;\qquad {\dot g}_2 = 4 g_1 g_2 + 2 g_4 g_5
\nonumber \\
{\dot g}_3 &=& g_5^2 + 5 g_4^2, \;\qquad 
{\dot g}_4 = g_5g_2 + g_4g_3 + 4 g_1 g_4
\nonumber \\
{\dot g}_5 &=& 5 g_4 g_2 + g_5 g_3 ,
\label{1RGhalfilling}
\end{eqnarray}
where ${\dot g}_a = \partial g_a/ \partial l$, $l$ being the RG time. The
resulting phase diagram is presented in Fig.~\ref{fig1}. As in two-leg
electronic ladders, there is a special isotropic ray of the RG flow
where an approximate SO(8) symmetry emerges in the far infrared limit
\cite{Lin}. Along the highly symmetric ray $g_a = g$ ($a=1,\ldots,5$),
model (\ref{majofthalfilling}) takes the form of the SO(8) Gross-Neveu
model which is an integrable massive field theory for $g>0$. The
resulting gapful phase is two-fold degenerate and corresponds to a
spin-Peierls (SP) ordering, with lattice order parameter ${\cal
  O}_{\rm SP} = \sum_{i, \alpha} (-1)^i [c^{\dagger}_{\alpha,i}
c_{\alpha,i+1} + \rm{H.c}]$. A second massive phase is obtained from
this SP phase by performing a duality transformation, $\xi^{7,8}_L
\rightarrow - \xi^{7,8}_L$, which is an exact symmetry of
Eq.~(\ref{majofthalfilling}) if $g_{4,5} \rightarrow - g_{4,5}$. This
duality symmetry exchanges a SP phase with a long-ranged charge
density-wave (CDW) phase with order parameter ${\cal O}_{\rm CDW}
= \sum_i (-1)^i \delta n_i$, where $\delta n_{i} = n_{i} -
\langle{n_{i}}\rangle$. The quantum phase transition between the
SP-CDW phases is found to belong to the U(1) universality class. There
is a second duality symmetry with $\xi^{6}_L \rightarrow - \xi^{6}_L$
which is a symmetry of Eq.~(\ref{majofthalfilling}) if $g_{2,5}
\rightarrow - g_{2,5}$. This duality symmetry is non-local in terms of
the original lattice fermions $c_{\alpha,i}$ and gives rise to two
non-degenerate fully gapped phases from SP and CDW phases. As it is
seen in Fig.~\ref{fig1}, a first non-degenerate phase contains the $V
< 0$ axis. Its physical interpretation is a singlet-pairing phase
which is the analog of the rung-singlet (RS) phase of the two-leg
ladder. Upon doping, the singlet BCS pairing $P_{00,i}$ has a gapless
behavior and becomes the dominant instability~\cite{sylvainMS}. We
need to introduce non-local string order parameters to fully
characterize the last non-degenerate phase. In this respect, we define
two charge string order parameters: $ {\cal O}^{\rm even}_{c,i} = \cos
\big( \frac{\pi}{2} \sum_{k < i} \delta n_{k} \big), \; {\cal O}^{\rm
  odd}_{c,i} = \delta n_{i} {\cal O}^{\rm even}_{c,i},$ which are
 even or odd, respectively , under the particle-hole transformation
$\delta n_{i} \rightarrow - \delta n_{i}$. Within the low-energy
approach, we find the long-range ordering of odd (even respectively)
charge-string operator in the second non-degenerate (RS respectively)
phase. The phase with long-range ordering of ${\cal O}^{\rm odd}_{c}$
is a HI phase similar to the Haldane phase of the spin-1
chain. Indeed, for attractive interactions $U,V < 0$, on general
grounds, we expect that the SO(5) spin gap ($\Delta_s$) will be the
largest scale of the problem. At energies lower than $\Delta_s$, one
can integrate out the SO(5) spin-degrees of freedom and the leading
part of the effective Hamiltonian (\ref{majofthalfilling}) simplifies
as follows:
\begin{equation}
{\cal H}_{\rm int} =
- i m_c \sum_{a=7}^{8} \xi_R^{a} \xi_L^{a}
- i m_o  \; \xi_R^{6} \xi_L^{6} ,
\label{majocharge}
\end{equation}
which is the well-known Majorana effective field theory of the spin-1
$XXZ$ Heisenberg chain with a single-ion anisotropy $D$
\cite{tsvelikspin1}. Along the special line $V = 2U$, the two masses
$m_{c,o}$ are equal due to the presence of the extended SU(2) symmetry
which rotates the three Majorana fermions $\xi^{6,7,8}$. Within the
spin-1 terminology, the interpretation of the phases for $U < 0$ of
Fig.~\ref{fig1} reads as follows: the CDW phase is the N\'eel phase, the RS phase
is the large-$D$ singlet phase and the HI phase is the Haldane
phase. All the known quantum phase transitions in the spin-1 problem
are consistent with the findings of the RG approach of model
(\ref{majofthalfilling}) with a U(1) quantum criticality for the
HI-RS transition and an Ising transition between the CDW and HI
phases.  The HI phase of Fig.~\ref{fig1} is characterized by a
string-order ${\cal O}^{\rm odd}_c$ which reveals its hidden order.  
We can also investigate the possible existence of edge
states in the HI phase by considering a semi-infinite geometry. In
that case, the low-energy effective Hamiltonian is still given by
Eq.~(\ref{majocharge}) with the boundary conditions: $\xi^{6,7,8}_{L}
\left(0\right) = \xi^{6,7,8}_{R} \left(0\right)$.  The situation at
hand is very similar to the low-energy approach of the cut two-leg
spin ladder \cite{orignac}. The resulting boundary model is integrable
and three localized Majorana modes ${\vec \eta}$ with zero energy
inside the gap (midgap states) emerge in the HI phase. These three
local fermionic modes give rise to a local pseudo spin-1/2 operator
$\vec{\mathscr{S}}$ thanks to the identity \cite{tsvelikmajo}:
$\vec{\mathscr{S}} = - i \; {\vec \eta} \wedge {\vec \eta}/2$. We thus
conclude on the existence of a spin-singlet pseudo-spin-1/2 edge state
which is the main signature of the HI phase.
 
\begin{figure}[t]
\centering
\includegraphics[width=0.95\columnwidth,clip]{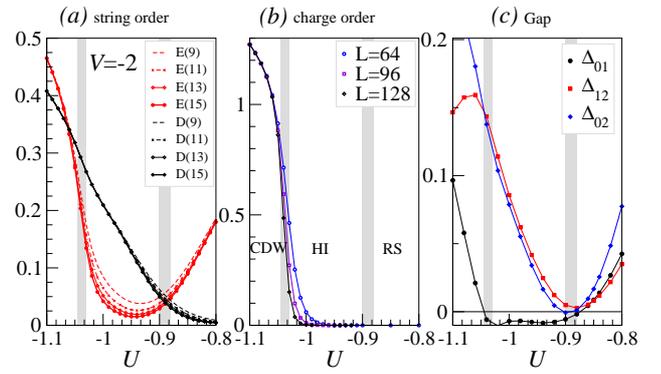}
\caption{(Color online): (a-b) Order parameters along the $V=-2$ line
  showing the three different phases CDW, HI and RS. String orders are
  computed by taking $i$ and $j$ at equal distance from the center of
  the chain. (c) Various charge gaps $\Delta_{ab}$ (see text),
  extrapolated in the thermodynamic limit from data obtained on
  $L=16$, $32$ and $64$. Quantum phase transitions are located in the
  grey regions.}
\label{fig:orders}
\end{figure}

\section{DMRG calculations} 

We now carry out numerical calculations,
using DMRG, in order to validate this conjecture in the $N=2$ and
$N=3$ cases. We keep up to 2000 states and use open boundary
conditions. When $N=2$, we fix two quantum numbers for the spin part
$S^z = \sum_{\alpha,i} (-)^{\alpha+1} n_{\alpha, i}/2$ and $T^z =
\sum_i(n_{1,i}+n_{2,i}-n_{3,i}-n_{4,i})/2$ (the ground state lies in
the $S^z=T^z=0$ sector), and the total number of particles $N_f = 2L$.
For $N=2$, we set $t=1$, $V = -2$ and we investigate order parameters
showing the existence of the HI phase and its extension.  In this
respect, we define two string order correlations: $E(|i-j|) =
|\langle{\exp\big(i\pi \sum_{i < k < j} \frac{\delta n_k}{2}\big)
}\rangle |$ and $D(|i-j|) = |\langle{ \frac{\delta n_i}{2} \exp\big(i
  \pi \sum_{i < k < j} \frac{\delta n_k}{2}\big) \frac{\delta
    n_j}{2}}\rangle|$.  In Fig.~\ref{fig:orders}, we plot these string
correlations, the charge order parameter $\langle{{\cal O}_{\rm
    CDW}(L/2)}\rangle$ in the bulk of the chain, and the pseudo-spin
gaps which are defined by:
\begin{equation*}
\Delta_{ab} = E_0(N_f=2L+2b) - E_0(N_f=2L+2a)\;,
\end{equation*} 
with $E_0(N_f)$ as the ground-state energy with $N_f$ particles and
$S^z=T^z=0$.  In the HI phase, because of the existence of edge states
(see below), the excited state with $N_f+2$ fermions falls onto the
ground-state (i.e. $\Delta_{01}=0$), so that the correct value for the
gap in the bulk is given by $\Delta_{12}=\Delta_{02}$, similarly to
what has been done for spin-1 chains.  All these quantities lead to
the conclusion of the existence of two gapful phases on top of the CDW
phase.  In particular, the data confirm the existence of the HI gapped
phase with $\langle{\cal O}_{\rm CDW}\rangle=0$, $D(\infty)\neq 0 $
while $E(\infty)$ scales to zero. On the contrary, $D(\infty) = 0$ in
the RS phase while $E(\infty)$ remains finite. In the CDW phase, both
string orders are finite, which can be easily understood from the
ground-state structure with alternating empty and fully occupied
sites.
\begin{figure}[t]
\centering
\includegraphics[width=0.66\columnwidth,clip]{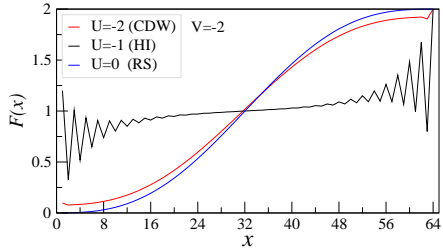}
\caption{(Color online) Integrated excess density $F(x)$ showing the
   edge states in the HI phase of the attractive model with
  $N=2$.}
\label{fig:edgestates}
\end{figure}
One of the striking features of the HI phase are the edge states. As
discussed above, these edge states are in the charge sector so one can
observe them by adding two particles while staying in the $S^z=T^z=0$
sector.  In Fig.~\ref{fig:edgestates}, we plot the integrated ``excess
density'' defined as $F(x) = \int^x\! dy (n(y)-2)$ for $U > -1$ and
$F(x) = (-1)^{x} \int^x \!dy (n(y)-2)$ if $U\leq -1$ to remove the
typical CDW oscillations. We find that, in the HI phase, the added
particles are pinned at the ends of the chains while in the RS and CDW
phases, this excess lies in the bulk.

\begin{figure}[t]
\centering
\vspace{1cm}
\includegraphics[width=0.9\columnwidth,clip]{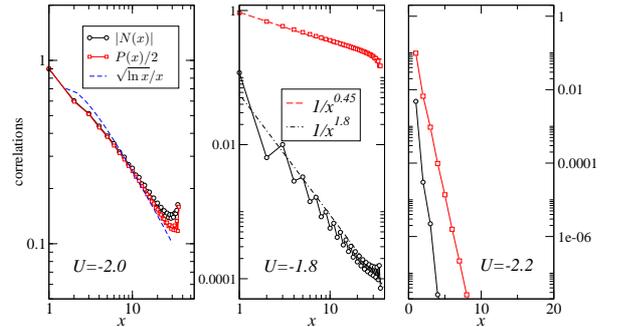}
\caption{(Color online) Correlation functions $|N(x)|$ and $P(x)$ for
  $N=3$, corresponding to the correlations of the ${\cal S}^z$ and
  ${\cal S}^{+}$ pseudo-spin operators, as a function of
  distance $x$ at $V=-6$ for $L=72$. For $V= 3 U$ (a), SU(2) symmetry is
  manifest, 
 while the system is XY-like if
  $U > V/3$ (b), or Ising-like if $U < V/3$ (c).}
\label{fig52}
\end{figure}

Finally, we discuss the case $N=3$, i.e. spin-5/2 fermions. As shown
in Fig.~\ref{fig52}, the system behaves \emph{effectively} as a
critical spin-$3/2$ SU(2) chain on the line $V= 3 U$, with equal
transverse and longitudinal pseudo-spin correlations given
 by the singlet-pairing $P(x)= \langle{
  P^{\dagger}_{00,L/2+x} P_{00,L/2}}\rangle$ and the charge
correlations $N(x)= \langle{\delta n_{L/2+x} \delta n_{L/2}}\rangle$, respectively.
In particular, we recover the same quantum critical behavior as a
spin-1/2 chain as predicted~\cite{haldane}.  
Moving away
from this line, we find in Fig.~\ref{fig52} the emergence of a
Luttinger liquid phase with critical exponents close to the one of the
$XY$ model for $U\geq V/3$, and a gapped Ising phase with exponentially
decaying correlations when $U<V/3$, in full agreement with the
strong-coupling approach.

The authors would like to thank J.~Almeida, P.~Azaria, M.~Fabrizio,
A.~A.~Nersesyan, and K.~Totsuka for illuminating discussions. P.~L. is
very grateful to the Abdus Salam ICTP for hospitality during the
completion of this work.


\begin{thebibliography}{99}
\bibitem{footnote}
Other on-site states acquire a large energy in this limit and can be
discarded.
\end{thebibliography}
\end{document}